\documentclass[9pt,twocolumn,twoside]{osajnl}

\journal{pr} 

\setboolean{shortarticle}{false}

\usepackage{lineno}
\usepackage{xcolor}

\title{Optical microcombs in whispering gallery mode crystalline resonators with dispersive intermode interactions}

\author[1]{Tuo Liu}
\author[1]{Suwan Sun}
\author[1]{You Gao}
\author[1]{Siyu Wang}
\author[1]{Yongyuan Chu}
\author[1,*]{Hairun Guo}

\affil[1]{Key Laboratory of Specialty Fiber Optics and Optical Access Networks, Joint International Research Laboratory of Specialty Fiber Optics and Advanced Communication, Shanghai University, Shanghai 200444, China}
\affil[*]{corresponding author: hairun.guo@shu.edu.cn}

\begin{abstract}
Soliton microcombs have shown great potential in a variety of applications ranging from chip scale frequency metrology to optical communications and photonic data center, in which light coupling among cavity transverse modes, termed as intermode interactions, are long-existing and usually give rise to localized impacts on the soliton state. 
Of particular interest are whispering gallery mode based crystalline resonators, which with dense mode families, potentially feature interactions of all kind.
While effects of narrow-band interactions such as spectral power spikes have been well recognized in crystalline resonators, that of broadband interactions remains unexplored.
Here, we demonstrate microcombs with broadband and dispersive intermode interactions, in home-developed magnesium fluoride microresonators with an intrinsic $\mathbf{Q}$-factor approaching 10 billion.
In addition to conventional soliton comb generation in the single mode pumping scheme, comb states with broadband spectral tailoring effect have been observed, via an intermode pumping scheme.
Remarkably, footprints of both constructive and destructive interference on the comb spectrum have been observed, which as confirmed by simulations, are connected to the dispersive effects of the coupled mode family.
Our results not only contribute to the understanding of dissipative soliton dynamics in multi-mode or coupled resonator systems, but also extend the access to stable soliton combs in crystalline microresonators where mode control and dispersion engineering are usually challenging.
\end{abstract}

\setboolean{displaycopyright}{true}

\begin{document}

\maketitle

\section{Introduction}
\label{sec-1}
Soliton microcombs based on optical microresonators have triggered the rapid development of miniature and chip scale optical frequency combs in recent years \cite{herr_temporal_2014, brasch_photonic_2016, yi_soliton_2015,  kippenberg_tobias_j_dissipative_2018}, and have resulted in the emergence of fully integrated frequency comb chips, opening an access to high-performance, high-compactness, and high-volume laser sources for advanced optical metrology \cite{stern_battery-operated_2018, raja_electrically_2019, shen_integrated_2020, xiang_chao_laser_2021, chang_integrated_2022}.
Indeed, a number of proof-of-concept applications have been demonstrated with soliton microcombs, such as massive parallel optical communications \cite{marin-palomo_microresonator-based_2017, fujii_dissipative_2022}, optical ranging \cite{suh_soliton_2018, trocha_ultrafast_2018, wang_long-distance_2020}, massive parallel LIDAR \cite{riemensberger_massively_2020}, low-noise microwave synthesis \cite{liang_high_2015, lucas_ultralow-noise_2020, liu_photonic_2020}, astronomical spectral calibration \cite{suh_searching_2019, obrzud_microphotonic_2019}, and optical nuerophomic computing\cite{feldmann_parallel_2021, xu_11_2021}.

While recent focus is on photonic integrated platforms where wafer scale, high quality and highly nonlinear optical microresonators are accessible \cite{liu_high-yield_2021}, a parallel platform is whispering gallery mode (WGM) based crystalline resonators \cite{braginsky1989quality, grudinin_ultra_2006}. 
In particular, crystalline fluoride resonators could have a record-high finesse beyond ${10^{7}}$ \cite{savchenkov_optical_2007}, which is suitable for the generation of ultra-narrow linewidth lasers \cite{sprenger_caf2_2010, alnis_thermal-noise-limited_2011, liang_ultralow_2015, lim_chasing_2017} as well as soliton microcombs\cite{herr_temporal_2014, pavlov_soliton_2017, liu_low-loss_2018, pavlov_narrow-linewidth_2018, fujii_dissipative_2022}. 
Given a weak thermo-refractive noise, such resonators could support solitons with low-noise repetition frequencies, serving as a photonic microwave synthesizer \cite{liang_high_2015, lucas_ultralow-noise_2020}.
Moreover, on the study of soliton physics, they represent an ideal platform which is almost free from high-order dispersive or nonlinear effects. 
A number of dissipative soliton dynamics have been demonstrated with high-level of agreement with the theory, such as the soliton double resonance \cite{guo_universal_2016}, soliton pulse scaling \cite{lucas_detuning-dependent_2017}, soliton breathers \cite{lucas_breathing_2017}, soliton molecules \cite{weng_heteronuclear_2020}, soliton crystals \cite{taheri_all-optical_2022} and soliton behaviors to intermode interactions \cite{guo_intermode_2017}.

\begin{figure*}[t!]
    \centering
    \includegraphics[width=1 \linewidth]{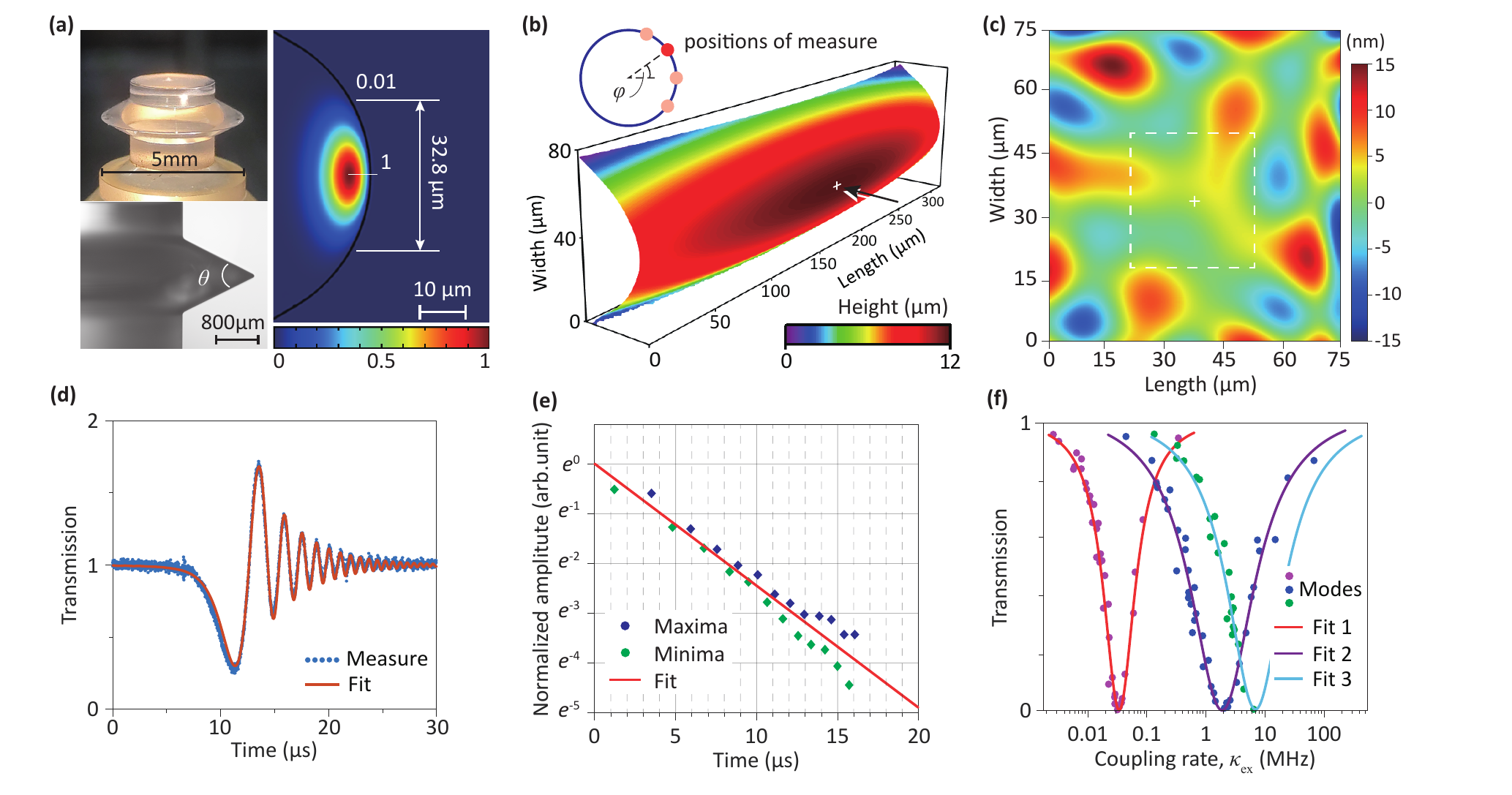}
    \caption{
    \textbf{Whispering gallery mode fluoride resonators.}
    (a) Microscopic pictures of a home-made ${\rm MgF_2}$ microresonator, in which a fundamental whispering gallery mode is numerically calculated and illustrated in the normalized form. The absolute amplitude at the peak of the mode field is normalized to unity. The positions on the cavity edge where the mode field amplitude is 1\% of the peak are marked, which reveals an axial range of ca. ${32.8 ~{\mu \rm m}}$ covering most of the light power in the cavity. 
    (b) A fractional 3D surface profile of the resonator, at a selected measure position on the edge. The colorbar indicates the radial height of the profile. 
    (c) The residual of the surface after being fitted and subtracted with a polynomial surface function, which in the area of ${75 \times 75 {\mu \rm m}^2}$ (where the fit shows a high level of approximation) is within ${\pm 15 ~\rm nm}$, and in ${32.8 \times 32.8 {\mu \rm m}^2}$ (corresponding to the 1\% mode field boundary) is within ${\pm 5 ~\rm nm}$. The overall root mean square of the extracted residual is ${3.5 ~\rm nm}$. 
    (d) The measured transmitted power trace of one resonance of a ${\rm MgF_2}$ resonator, which shows a ring-down profile at the ending edge upon a laser tuning speed of  ${350 ~{\rm GHz/s}}$. Fitting of this resonance reveals: ${Q_0 = 8.44 \times 10^9}$, ${Q_e = 1.1 \times 10^{10}}$, and the loaded ${Q}$-factor is ${Q = 4.75 \times 10^{9}}$.
    (e) Extracted maxima and minima from the resonance trace in (a), as the function of time, which reveal a decreased linear proportion of the photon decay rate over time. The fitted photon life time (${e^{-1}}$ level) is ${4.08 ~{\mu \rm s}}$, corresponding to a loaded ${Q}$-factor of ${4.95 \times 10^{9}}$.
    (f) Assessment of the ideality of three resonant modes in the ${\rm MgF_2}$ resonator. The plot shows the distribution of the central transmission of these resonances, upon tuning the external coupling rate.
    }
    \label{F1}
\end{figure*}

\begin{figure*}[t!]
    \centering
    \includegraphics[width=1 \linewidth]{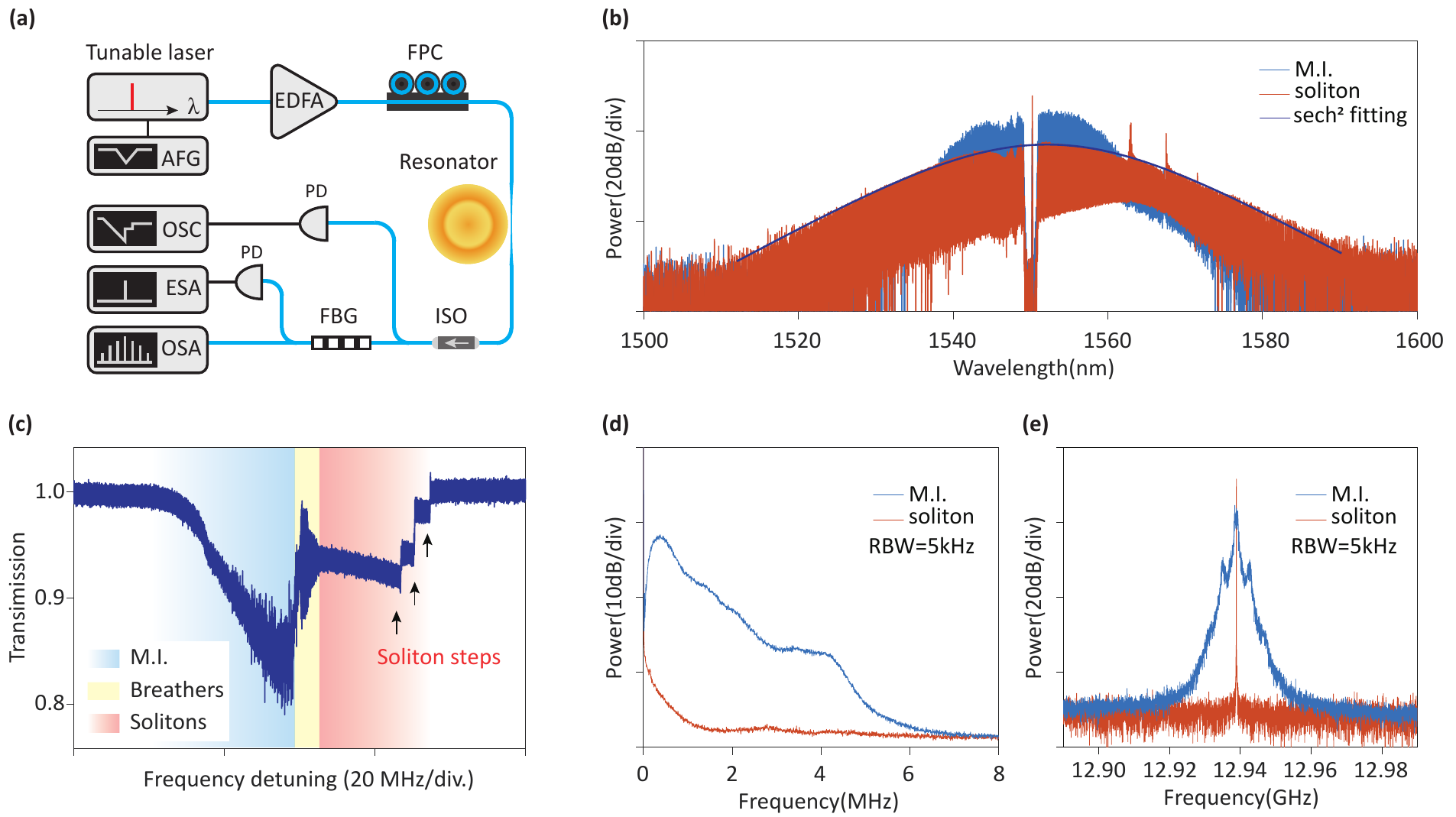}
    \caption{
    \textbf{Cavity \emph{Q}-factors and soliton comb generation.}
    (a) A schematic of the experimental setup. AFG: arbitrary function generator, EDFA: Erbium-doped fiber amplifier, OSC: oscilloscope, ESA: electrical spectrum analyzer, OSA: Optical spectrum analyzer.
    (b) The spectra of both an MI comb and a soliton comb.
    (c) The transmission of the resonance that is pumped and scanned by an intense cw laser ($186.3 ~{\rm mW}$), in which the stair-like pattern (soliton steps) indicates the formation of dissipative solitons in the cavity.
    (d) The low frequency \emph{rf} spectrum of the filtered combs, indicating the noisy and the low-noise nature of the MI and the soliton comb states, respectively.
    (e) The \emph{rf} spectrum around the repeat frequency of the soliton comb, which shows a narrow linewidth single frequency tone with signal to noise ratio (SNR) >60dB. The linewidth is measured to be $\mathcal{O}(10~\rm{kHz})$. The spectrum for the MI comb is also presented, which shows a broad line-shape. 
    }
    \label{F2}
\end{figure*}

Indeed, effects of intermode interactions have been widely observed and studied in soliton microcombs, which in most cases are reflected as localized power enhancement on certain comb modes \cite{herr_mode_2014}. 
These modes were also referred as single-mode dispersive waves, which would help to stabilize the soliton state if located in a particular "quiet point" \cite{yi_single-mode_2017}. 
In the strong coupling condition, energy exchange between mode families was also observed, which leads to instability on the soliton state, including a breathing-like dynamic \cite{guo_intermode_2017}.
Efforts were also made to engineer intermode interactions in microresonators. 
In particular in coupled resonator systems, lithographically engineering the coupling strength and the coupling frequency has evoked new features of cavity solitons, including the soliton comb with tunable dispersive waves \cite{tikan_emergent_2021}, dark pulse excitation with tailored dispersion profile \cite{xue_mode-locked_2015, jang_dynamics_2016, helgason_dissipative_2021}, and solitons with super efficiency \cite{xue_super-efficient_2019, helgason_power-efficient_2022}.

Recent studies have also shown that with an induced auxiliary mode, an intermode pumping wave may cause the thermal equilibrium in the cavity, and recover the access of soliton microcombs with reduced thermal instability \cite{li_stably_2017, weng_directly_2021,lei_thermal_2022}.

Usually, such coupled mode or coupled resonator systems require a dedicated design to apply proper intermode interactions.
In this regard, crystalline resonators with a high number of mode families would equally support a high volume of intermode interactions of different kind, and serve as an alternative platform in the study of emergent features of cavity solitons.
Nevertheless, most reported effects of intermode interactions in crystalline resonators were narrow-band and involving few comb modes, given a large difference in the free spectral range (FSR) between the coupled mode families. 
Effects of group velocity dispersion (GVD) in the second mode family, namely \emph{dispersive} intermode interactions, were usually hindered and unexplored.

\begin{table}[b!]
\centering
\caption{
Polishing slurry granule size, polishing time and corresponding roughness.
}
\begin{tabular}{cccc}
    \hline
    Granule size & Time & RMS Roughness & Q \\
    \hline
    1300 nm & 35-40 min& 125-160 nm &  $\rm 10^5$ \\
    250 nm & 20-30 min & 12-20 nm &$\rm 10^8$  \\
    50 nm & 15-20 min & 3-5 nm &$\rm 10^9$ \\
    \hline
   \end{tabular}
   \label{T1}
\end{table}

In this paper, we demonstrate soliton microcombs in crystalline magnesium fluoride ($\rm MgF_2$) resonators.
Effects of broadband and dispersive intermode interactions are experimentally evidenced, when an intermode pumping scheme is applied, which lead to tailored comb spectrum dependent on the dispersion in the second mode family. 
We also demonstrate a soliton comb in company with the Raman lasing, including both narrow-band lasers and Raman-Kerr combs \cite{chembo_spatiotemporal_2015, cherenkov_raman-kerr_2017, yu_raman_2020, xia_engineered_nodate}
A suspect Raman-soliton comb \cite{yang_stokes_2016, tan_multispecies_2021} is also observed around the anti-stokes mode.

\section{Results}
\label{sec-2}
\noindent\textbf{High-${\mathbf Q}$ WGM crystalline resonators}\\
First, we prepared ultra-high-${Q}$ crystalline ${\rm MgF_2}$ resonators, by means of mechanical machining\cite{coillet_microwave_2013,lin_barium_2014,fujii_all-precision-machining_2020}.
The preform of the cavity is diamond turned from a cylindrical bulk material, followed by surface polishing processes.
Diamond slurries with three particle sizes (i.e. 1300 nm, 250 nm, and 50 nm) are used in the polishing in an order to improve the surface smoothness, see Tab. \ref{T1}.
During the polishing, the surface profile as well as the roughness is also monitored by a commercial optical interferometric profiler (Sensofar Metrology, system's noise ${<1 ~{\rm nm}}$).
Figure \ref{F1}(a,b,c) showcase a measure of the cavity surface.
In particular, by having the detailed geometric structure, the transverse mode field of cavity's WGMs is numerically calculated, by which an axial range of care is defined to cover an area where the absolute mode field amplitude is greater than 1\% of its central peak, see Fig. \ref{F1}(a).
The measured profile is also fitted by a 2D polynomial function, and the surface roughness is extracted as the residual of the fit, see. Fig. \ref{F1}(b,c).
Remarkably, the estimated RMS roughness can be as low as a few nano-meters, which indicates a high quality surface polishing and an ultra-high-${Q}$ factor (typ. $>10^9$) of the resonator.
In practice, such a surface assessment is carried out at 18 positions along the circumference of the resonator (spaced by every 20 degree), indicating a variation of 5\% on the cavity diameter (i.e. 0.25 mm over a diameter of ca. 5 mm).
The curvature on the edge of the cross-section is ca. ${35 ~{\mu \rm m}}$.

The ${\rm MgF_2}$ resonator is then coupled with a tapered fiber for the assessment of the ${Q}$-factor and for soliton microcomb generation, using an experimental setup shown in Fig. \ref{F2}(a).
The transmission of the cavity resonance is measured by scanning the cw diode laser over it.
For ultra-high-${Q}$ resonators, the resonance linewidth is already comparable with that of the probe laser (${{\mathcal O}(10 ~{\rm kHz})}$), and the laser frequency tuning within the linewidth could be faster than the intracavity photon lifetime. 
This leaves a ring-down profile at the ending edge of the resonance, see Fig. \ref{F1}(d), which is also theoretically derived in Ref \cite{dumeige_determination_2008}.
In detail, the time evolution of the intracavity field of one resonant mode could be described as a a simple harmonic oscillator model, i.e.:
\begin{equation}
    \frac{\partial A}{\partial t} = i \omega_0 A- \frac{\kappa}{2} A + \sqrt{\kappa_{\rm ex}} \cdot s_{\rm in}
\label{E1}
\end{equation}
where $\omega_0$ is the angular frequency of the resonance, ${\kappa}$ is the overall loss rate of the system which consists of both the intrinsic loss rate ${\kappa_0}$ and the coupling loss rate ${\kappa_{\rm ex}}$, i.e. ${\kappa = \kappa_0 + \kappa_{\rm ex}}$, ${s_{\rm in}}$ stands for the external source, which in the frequency tunable cw mode can have the variation ${s_{\rm in} = s_0 \cdot \exp{(i\varphi(t))}}$. In the stationary form, i.e. the laser frequency tuning with respect to its initial value $\omega_i$ is slowly changed compared with the intracavity photon lifetime, we have ${\varphi(t) = \omega_i t }$, and the solution of Eq. \ref{E1} gives a standard Lorentz profile as a function of the laser detuning ${\delta_{\omega} = \omega_0 - \omega_i}$. In contrast, in the condition that the laser frequency tuning is comparable with the photon lifetime, we have ${\varphi(t) = (\omega_i + \frac{V_{\rm s}t}{2}) t}$, where ${V_{\rm s}}$ is the tuning speed of the laser frequency. Hence, a variation of Eq. \ref{E1} could be:
\begin{equation}
    \frac{\partial a}{\partial t} = i (\delta_\omega - V_{\rm s}t) a - \frac{\kappa}{2} a + \sqrt{\kappa_{\rm ex}} \cdot s_0
\label{E2}
\end{equation}
where ${A = a \cdot \exp{(i\varphi(t))}}$. The integration of Eq. \ref{E2} further gives:
\begin{equation}
    a = \sqrt{\kappa_{\rm ex}} \cdot s_0 \cdot \exp{\left(i \delta_\omega t - \frac{\kappa}{2}t \right)} \left[ f(t) - \frac{1}{i \delta_\omega - \frac{\kappa}{2}} \right]
\label{E3}
\end{equation}
with
\begin{multline*}
    f(t) = -\sqrt{\frac{i\pi}{2 V_s}} \exp{\left( i\frac{(i\delta_\omega-\kappa/2)^2}{2 V_s} \right)} \\ 
    \times \left[ {\rm erf} \left(-i\frac{i(\delta_\omega - V_{\rm s} t) - \kappa/2}{\sqrt{2 i V_{\rm s}}} \right)- {\rm erf} \left( -i\frac{i\delta_\omega - \kappa/2}{\sqrt{2 i V_{\rm s}}} \right) \right]
\end{multline*}
where ${\rm erf}(z)$ is the complex error function, i.e. ${z \in \mathbb{C}}$. The cavity transmission is then calculated as:
\begin{equation}
    T = \left| \frac{s_0-\sqrt{\kappa_{\rm ex}} \cdot a}{s_0} \right|^2
\label{E4}
\end{equation}

Note that fitting with Eq. \ref{E4} would deterministically return the coupling rate ${\kappa_{\rm ex}}$ as part of the overall loss rate $\kappa$, such that the coupling regime of the cavity (over- or under-coupling) can be distinguished.
As a result, an intrinsic ${Q}$-factor ${Q_0 = \omega_0/\kappa_0 = 8.44 \times 10^9}$ is obtained with respect to the transmission profile shown in Fig. \ref{F1}(d).

In contrast to the theoretical fitting, reading out the decay rate of the ring-down profile would also reveal the photon lifetime of the resonator, as well as the overall loaded ${Q}$-factor (i.e. ${Q = \omega_0/\kappa}$) \cite{savchenkov_optical_2007,lecaplain_mid-infrared_2016}, see Fig. \ref{F1}(e).
As a comparison, the loaded ${Q}$-factor via the theoretical fitting is ${4.75 \times 10^9}$ and that from the decay rate is ${4.95 \times 10^9}$.

Therefore, we have demonstrate ultra-high-${Q}$ ${\rm MgF_2}$ resonators, with an convincing ${Q}$-factor above ${10^9}$.
In addition, the coupling ideality with respect to selected resonant mode of the resonator is characterized, see Fig. \ref{F1}(f), which is on a high-level indicating almost no loss at the coupling junction \cite{spillane_ideality_2003, pfeiffer_coupling_2017}.

\begin{figure*}[t!]
    \centering
    \includegraphics[width=1
    \linewidth]{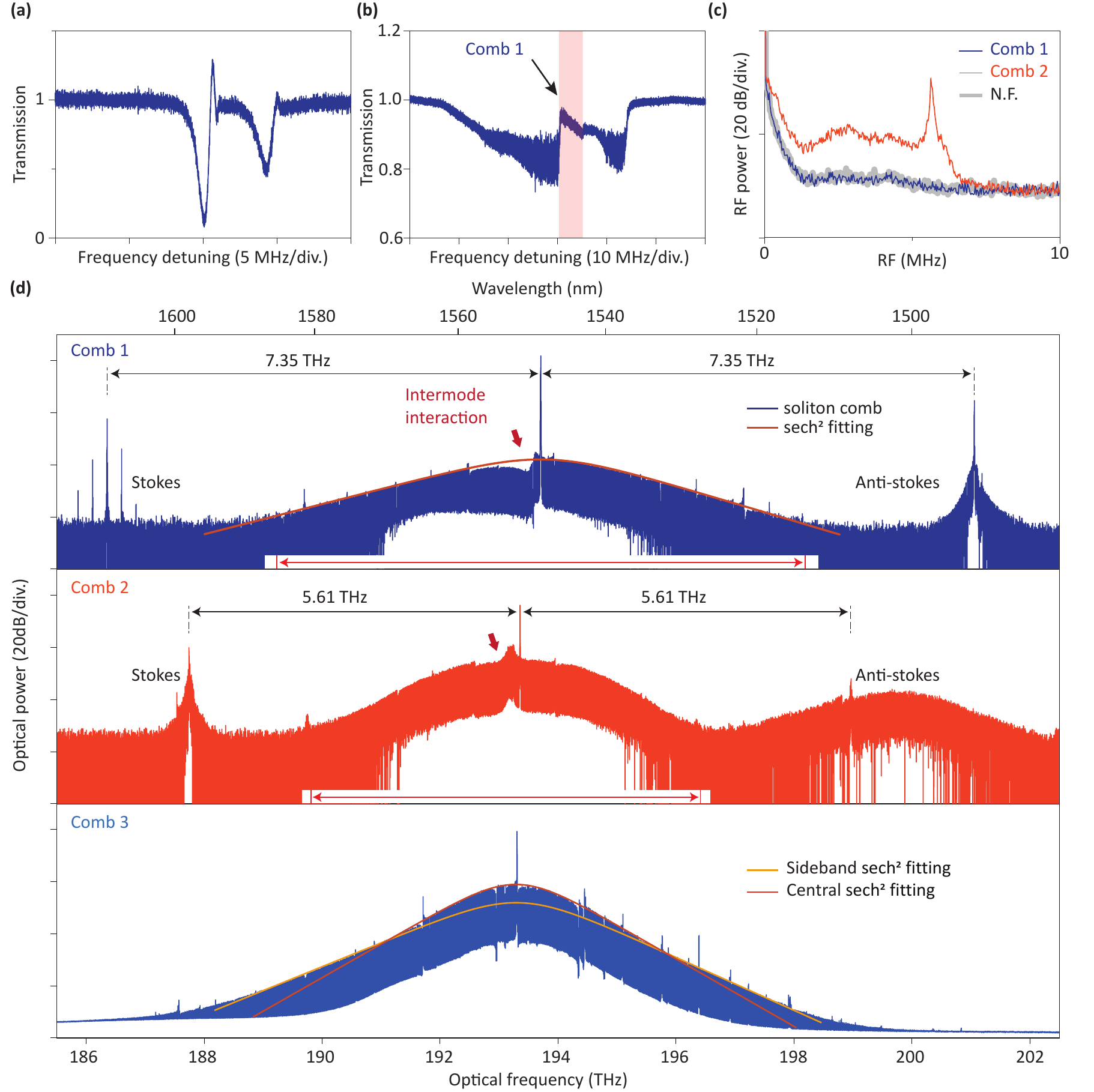}
    \caption{
    \textbf{Microcombs generated with the intermode pumping scheme.}
    (a) Measured transmission of two resonances in proximity, probed by the laser at low power (ca. 3.8 mW).
    (b) The transmission of the same resonances probed at high power (ca. 281.7 mW), which reveals a soliton step.
    (c) Low noise \emph{rf} spectrum of two combs (comb 1 and comb 2 shown in (d)) via the intermode pumping.
    (d) The measured comb spectra. Comb 1 is generated corresponding to the soliton step region in (b), while comb 2 is pumped regarding a second pair of resonances and is in a changed polarization state. Both combs feature intermode interactions on the central portion, and strong Raman lasing at both stokes and anti-stokes bands. In particular, a suspect anti-stokes soliton comb are observed in comb 2. The full transmitted power is 238.2 mW (comb 1) and 197.3 mW (comb 2), respectively. The central comb power is estimated within the marked range (red arrows) such that the power of the Raman lasing is excluded in the calculation of power efficiency.
     Comb 3 is another observation that features broadband central power enhancement and the overall spectral envelope is beyond the standard ${\rm sech^2}$-profile (compared with spectral fittings on both the central portion and on the sideband of the comb). As a consequence, the power efficiency of this comb state is also increased (pumping power ${ 192 ~{\rm mW}}$ and the transmitted comb power ${ 133 ~{\rm mW}}$.
    }
\label{F3}
\end{figure*}

\begin{figure*}[t!]
    \centering
    \includegraphics[width=0.7 \linewidth]{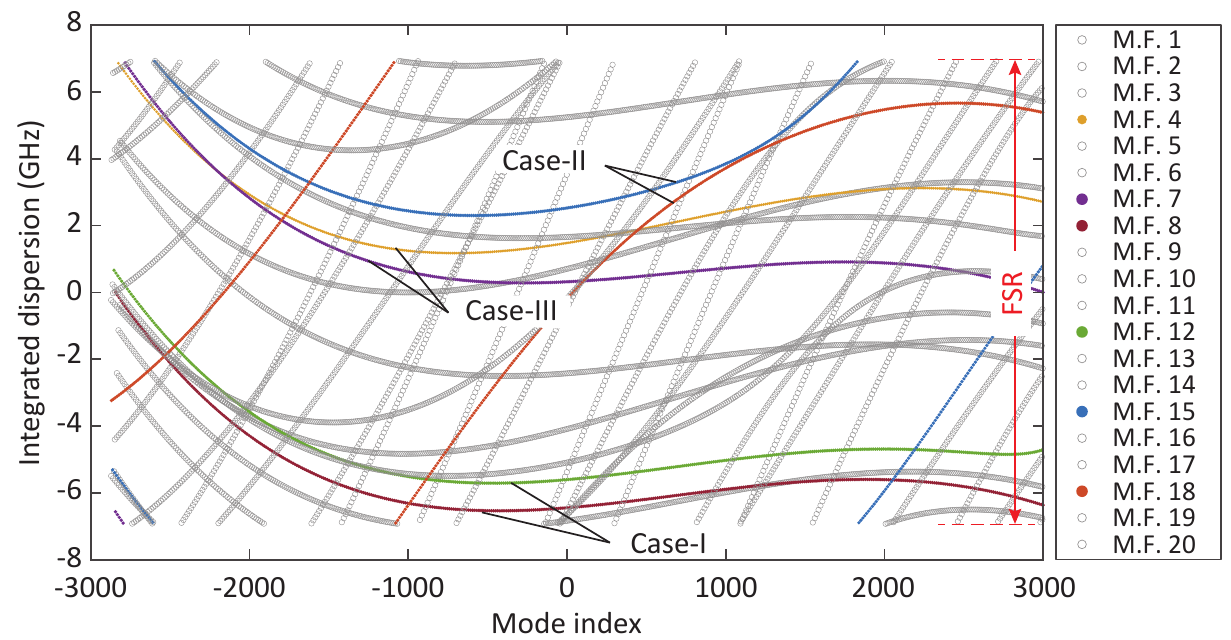}
    \caption{
    \textbf{simulated mode map of 20 transverse mode families in magnesium fluoride resonators.}
    For each mode family (M.F.), the resonant frequencies are compared to a uniform frequency grid $\omega_0 + \mu D_{1}$ (where $\omega_0/2\pi$ and $D_{1}/2\pi$ are set as the central mode frequency and the FSR of the mode family 1).
    Thus the relative frequency difference among resonant modes is extracted and illustrated.
    In the map, six mode families are highlighted, showing three pairs of dispersion relations (marked as case-I, -II and -III).
    }
\label{F5}
\end{figure*}

\vspace{3 mm}
\noindent\textbf{Soliton microcomb generation}\\
We next carried out experiments for soliton microcomb generation in ${\rm MgF_2}$ resonators.
By conventional laser tuning scheme with respect to an isolated high-$Q$ cavity resonance, we could reproduce the classic transmittance of the system as a function of the laser-cavity detuning, which reveals soliton steps on the red-detuned side of the resonance\cite{herr_temporal_2014}, see Fig. \ref{F2}(c).
In addition, the detuning region corresponding to the modulation instability (MI) regime and to the soliton breathers are also noted.
Tuning the laser from the blue-detuned side and landing on the soliton step, we could observe soliton microcomb generation, especially the single soliton state that has an overall smooth $\rm sech^2$-shape spectral envelope, see Fig. \ref{F2}(b).
Moreover, the stability of the soliton comb is assessed by measuring its low-frequency \emph{rf} spectrum (Fig. \ref{F2}(d)) as well as the repetition frequency (Fig. \ref{F2}(e)), after the residual pump wave was removed by a fiber notch filter before the spectrum was photo-detected. 
As a result, while the comb in the MI regime shows a noise figure around DC and a broad line shape around the repetition frequency, the soliton comb shows a low noise state with a narrow linewidth repetition frequency.
The linewidth of the soliton repetition frequency is measured to be ${\mathcal{O}(10~\rm{kHz})}$, similar to that of the pumping wave.

\vspace{3 mm}
\noindent\textbf{Microcombs with broadband intermode interactions}\\
In fact, there is a large number of mode families in our crystalline resonators, each corresponding to a transverse eigen-mode field (in the ($r,z$) plane).
As a consequence, there is a high probability to observe resonances that are in proximity, such that the intermode pumping scheme can be investigated.
Figure \ref{F3}(a) showcases a transmission of two resonances in proximity, via the laser tuning at the power of $3.8 ~{\rm mW}$. 
When the pump power is increased to $\sim 280 ~{\rm mW}$, which is sufficient to excite the comb generation, a complex transmittance of the system is detected, see Fig. \ref{F3}(b), and a soliton step situated in between the two resonances is observed. 
Indeed, tuning the laser frequency from the blue detuned side to stop on the soliton step, a soliton comb spectrum is observed, see comb 1 in Fig. \ref{F3}(d).
Intuitively, the soliton state is stemming from pumping the primary (left) resonance and is on the slope of the auxiliary (right) resonance, such that it may feature interactions in between.
As previously reported, one effect is that the pumping wave being coupled to the second resonant mode would cause the thermal equilibrium of the system \cite{weng_directly_2021,lei_thermal_2022}.
This effect also applies to our system, where we noted that the soliton comb would be running for hours free from additional feedback control (despite that the laser drift would surpass the soliton existence range during the time). 

Moreover, we noted that the generated comb spectrum via the intermode pumping scheme usually features a strongly tailored spectral envelope. 
This includes not only two comb spectra (i.e. comb 1 and comb 2 in Fig. \ref{F3}(d)) featuring certain spectral profiles at the center, but also a comb spectrum (comb 3) with broadband enhancement such that the overall envelope is beyond the standard ${\rm sech^2}$ profile.



In addition, the Raman lasing is observed in our ${\rm MgF_2}$ resonators, in company with the microcomb generation, see Fig. \ref{F3}(d).
The noted frequency shift for both the stokes and anti-stokes modes is ca. 7.35 THz (comb 1), and ca. 5.61 THz (comb 2 with a different polarization), which are much smaller compared with that of reported Raman bonds of fluoride materials, and are attributed to Raman bonds of isolated ${\rm MgF_2}$ oligomers on the surface of the resonator \cite{neelamraju_experimental_2012}.
We observed not only isolated Raman lasing and narrow-band Raman-Kerr combs as usual states \cite{chembo_spatiotemporal_2015}, but also a suspect Raman-soliton comb \cite{yang_stokes_2016} on the anti-stokes band, which has a wide spectral span and has a similar ${\rm sech^2}$-shape envelope to conventional soliton microcombs.

The stability of microcombs via the intermode pumping is also assessed by measuring the low-frequency \emph{rf} spectrum, see Fig. \ref{F3}(c).
It can be noted that the soliton comb 1 accompanied with narrow-band Raman-Kerr combs could feature a low-noise figure, indicating that the comb is overall in a stable state.
The comb 2 features an outstanding \emph{rf} tone at ca. 6 MHz, together with certain noise, which indicates that the comb is in a transition from the MI to the breathing state.

We also calculated the conversion efficiency of the observed microcombs, using the following equation \cite{xue_microresonator_2017}: 

\begin{equation}
    \eta = \frac{\sum_\mu{P_\mu} - P_0}{{P_{\rm in}}}
\end{equation}
where $P_\mu$ indicates the power of the comb mode and $\mu$ is the mode index, $P_0$ is the power of the central mode which mostly contains the residual pump power, $P_{\rm in}$ is the pumping power in the tapered fiber before being coupled into the resonator.
As a result, the soliton comb generated with the conventional pumping scheme (Fig. \ref{F2}(b)) has an efficiency of 1.28\%.
The soliton comb 1 in Fig. \ref{F3}(d) has an efficiency of 0.99\% (taking the central comb) with certain pump power converted to the Raman lasing.
The efficiency for the comb 2 is 9.54\% (also taking the central comb), yet it's not a fully stabilized comb state.
The comb 3 is calculated to have an outstanding efficiency, ca. 31.1\%. 
However, there lacks of essential evidence to show if this comb is in the low noise or soliton state.

Physically, the efficiency of soliton combs in a single and anomalous mode family is governed by the cavity nonlinearity as well as affected by the coupling scheme, which in $\rm MgF_2$ microresonators, with the nonlinear coefficient ${\sim 10^{-4} ~{\rm Hz}}$ per photon and in the strongly over coupled regime, could only reach a few percentage.
As such, We notice that the efficiency of >30\% in the comb 3 is truly outstanding, and can be attributed to effects \emph{beyond} the single mode operation.
That is, the pump power partially coupled into the auxiliary mode family would via intermode interactions transfer back to the comb mode and increase the comb power, and with constructive interference, the power efficiency is increased.
In practice, microcombs with close to the 0-dBm power level (excluding the residual pump) would be highly desired applications including telecommunications \cite{marin-palomo_microresonator-based_2017,marin-palomo_performance_2020,fujii_dissipative_2022} and photonic microwave synthesis \cite{liang_high_2015, lucas_ultralow-noise_2020, liu_photonic_2020}.

\begin{figure*}[t!]
    \centering
    \includegraphics[width=1 \linewidth]{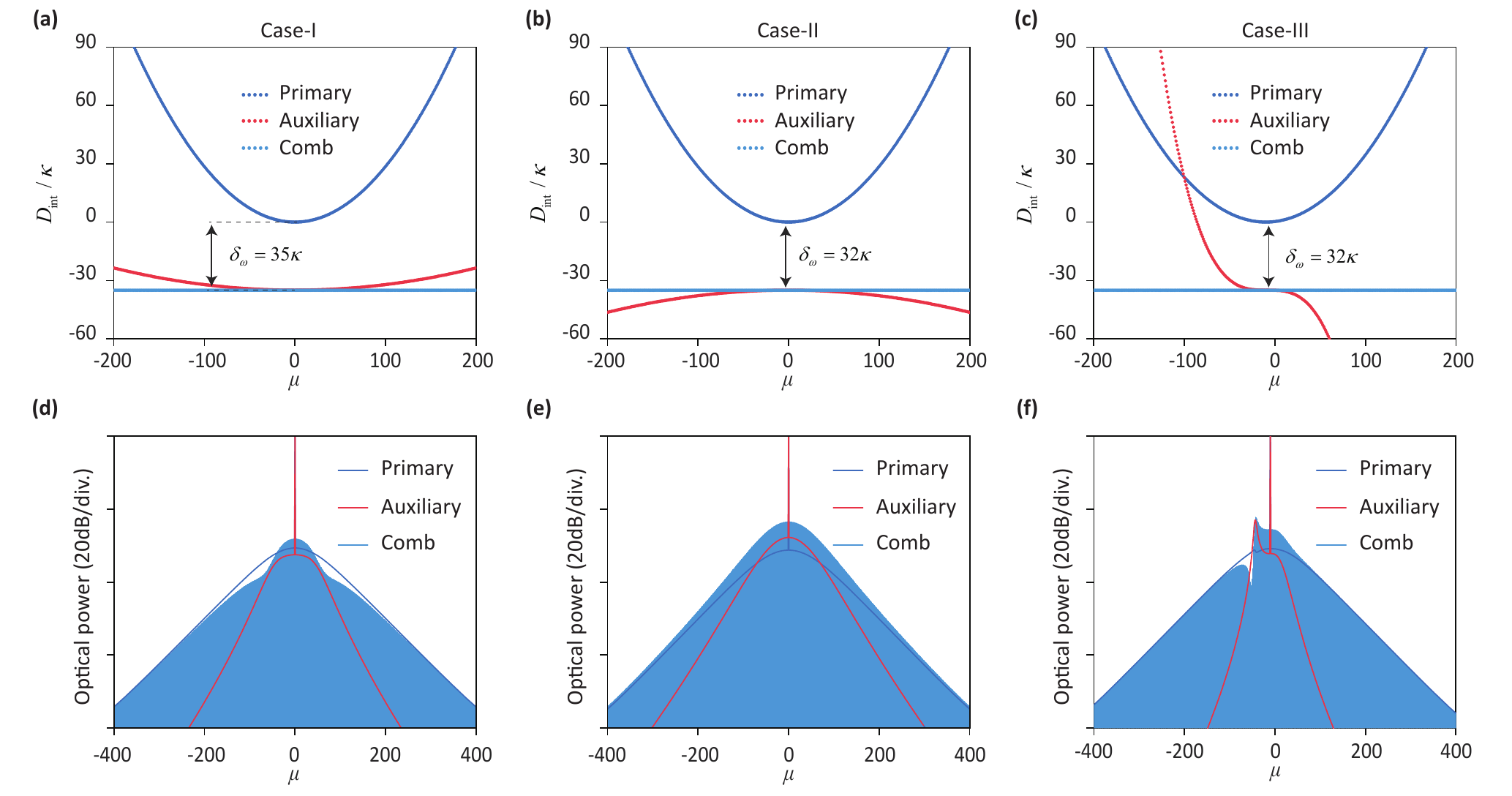}
    \caption{
    \textbf{Simulation of soliton combs with intermode interactions.}
    (a, b, c) Dispersion profiles of two coupled mode families, in which the primary mode family (blue dots) is unchanged and the auxiliary mode family (orange dots) shows three different profiles, corresponding to anomalous, normal and close-to-zero dispersion with respect to the central mode (${\mu = 0}$). The generated comb modes (light blue dots) at a certain detuning ($\delta_\omega$) is reflected as a straight line in these mode maps.
    (d, e, f) Transmitted spectra separately from the primary and the auxiliary mode families, i.e. $(1-\alpha)|s_{out,p}|^2$ (blue line) and $\alpha|s_{out,aux}|^2$ (orange line), respectively, and the combined spectrum $|s_{out}|^2$ (light blue drop lines) revealing wave interference in between.
    }
\label{F4}
\end{figure*}

\vspace{3 mm}
\noindent\textbf{Numerical simulations}\\
For a better understanding of the comb spectral tailoring effect with the intermode pumping, We referred to numerical simulations based on a pair of coupled Lugiato-Lefever equations (LLE) \cite{guo_intermode_2017}: 

\begin{equation}
\begin{split}
    \frac{\partial \tilde{A}_{\rm p}(\mu,t)}{\partial t} &=  \left(-\frac{\kappa_{\rm p}}{2} + i \delta_\omega + iD_{\rm int}(\mu)\right)\tilde{A}_{\rm p} + i\kappa_{\rm c} \tilde{A}_{\rm aux} \\
   & -ig_{\rm p} \mathcal{F}\left[|A_{\rm p}|^2 A_{\rm p}\right]_{\mu} -ig_{\rm c} \mathcal{F}\left[2|A_{\rm aux}|^2 A_{\rm p}\right]_{\mu}\\
    &+ \sqrt{\kappa_{\rm ex,p}}\sqrt{1-\alpha}s_{\rm in}
    \end{split}
\end{equation}

\begin{equation}
\begin{split}
    \frac{\partial \tilde{A}_{\rm aux}(\mu,t)}{\partial t} &= \left(-\frac{\kappa_{\rm aux}}{2} + iD_{\rm aux}(\mu)\right)\tilde{A}_{\rm aux} + i\kappa_{\rm c} \tilde{A}_{\rm p} \\
   & -ig_{\rm aux} \mathcal{F}\left[|A_{\rm aux}|^2 A_{\rm aux}\right]_{\mu} -ig_{\rm c} \mathcal{F}\left[2|A_{\rm p}|^2 A_{\rm aux}\right]_{\mu}\\
   & + \sqrt{\kappa_{\rm ex,aux}}\sqrt{\alpha}s_{\rm in}
     \end{split}
\end{equation}
where $|\tilde{A}_{\rm p}(\mu,t)|^2$ stands for the photon number in the primary mode family (where the soliton formation is expected), and $|\tilde{A}_{\rm aux}(\mu,t)|^2$ is that in the auxiliary mode family, $A_{\rm p}(\theta,t)$ and $A_{\rm aux}(\theta,t)$ are amplitudes in the Fourier domain with respect to the mode index $\mu$,
$\kappa_{\rm p}$ and $\kappa_{\rm aux}$ are overall loss rates of cavity mode families, including both the intrinsic loss rate and the coupling loss rate.
In particular, the coupling loss rates are expressed as $\kappa_{\rm ex,p}(1-\alpha)$ and $\kappa_{\rm ex,aux}\alpha$, where the subparameters ${\alpha}$, ${1-\alpha}$ are introduced revealing fundamental modal overlap coefficients between the bus waveguide mode (in the tapered fiber) and the cavity modes (and in a close system with two cavity modes,  the sum of the coefficients is forced to be unity), and subparameters $\kappa_{\rm ex,p}$ and $\kappa_{\rm ex,aux}$ are more specifically referred as the coupling strength determined by the gap distance between the fiber and the cavity.
This enables us to separate and monitor the transmitted light fields from the primary and the auxiliary mode families.
The coefficient $\kappa_{\rm c}$ is the linear coupling strength between the two mode families.
The laser-cavity detuning is $\delta_\omega = \omega_{\rm p}(0) - \omega_{\rm i}$, where $\omega_{\rm p}$ is the angular frequency of resonances in the primary mode family, and $\omega_{\rm i}$ is that of the pumping wave.
The resonant mode $\mu = 0$ is the central pumped mode and $s_{\rm in}$ is the pump amplitude.
The integrated dispersion of the primary mode family is $D_{\rm int}(\mu) = \omega_{\rm p}(\mu) - \omega_{\rm p}(0) - \mu D_{\rm 1,p}$, where $D_{\rm 1,p}/(2\pi)$ is the free spectral range (FSR) around the pumped mode.
In the same reference frame, the dispersion profile of the auxiliary mode family is $D_{\rm int}(\mu) = \omega_{\rm aux}(\mu) - \omega_{\rm p}(0) - \mu D_{\rm 1,p}$.
Nonlinear coupling coefficients $g_{\rm p}$ and $g_{\rm aux}$ indicate the single-photon induced nonlinear frequency shift by means of the self-phase modulation, and $g_{\rm c}$ is that by the cross-phase modulation.
Indeed, $g_{\rm p}(\mu) \propto n_2/V_{\rm eff,p}(\mu)$ and $g_{\rm c}(\mu) \propto n_2/V_{\rm eff,c}(\mu)$, where $n_2$ is the nonlinear refractive index of the material, $V_{\rm eff,p}(\mu)$ is the effective mode volume of the primary mode family and $V_{\rm eff,c}(\mu)$ is the effective overlapped mode volume between the primary and the auxiliary mode families.

Then, the transmitted field pattern of each mode family is:
\begin{equation}
    s_{\rm out,p} = \sqrt{1-\alpha}s_{\rm in} - \sqrt{\kappa_{\rm ex,p}}\tilde{A}_{\rm p}
\end{equation}
\begin{equation}
    s_{\rm out,aux} = \sqrt{\alpha}s_{\rm in} - \sqrt{\kappa_{\rm ex,aux}}\tilde{A}_{\rm aux}
\end{equation}
and the overall transmitted field is:
\begin{equation}
\begin{aligned}
   s_{\rm out} &= \sqrt{1-\alpha}s_{\rm out,p} + \sqrt{\alpha}s_{\rm out,aux} \\
              &= s_{\rm in} - \sqrt{\kappa_{\rm ex,p}}\sqrt{1-\alpha}\tilde{A}_{\rm p}-\sqrt{\kappa_{\rm ex,aux}}\sqrt{\alpha}\tilde{A}_{\rm aux}
\end{aligned}
\end{equation}
which contains wave interference between two field patterns, both transmitted and propagated through the tapered fiber. 

Before LLE simulations, we first performed mode simulations based on the finite element method (COMSOL). 
The simulations take the geometry shown in Fig. \ref{F1}(a), and cover 20 transverse mode families, and could confirm certain information of the ${\rm MgF_2}$ resonators, such as the FSR and the GVD of mode families. 
The simulated $D_2/2\pi$ component is between 1-2 kHz for most of the mode families, while few higher order mode families would feature normal group velocity dispersion.
Moreover, a mode map of all 20 mode families is illustrated, see Fig. \ref{F5}, which reveals a high number of mode crossings in the resonator.
Physically, a mode crossing indicates that there is phase matching between different mode families, which under perturbations (e.g. via imperfection induced light scattering in the resonator) would induce light coupling in between, efficiently at the crossing modes.
In the map, we noticed that some of the mode crossings are sharp such that the light coupling is expected among few modes, namely narrow-band intermode interactions.
However, there are also mode families featuring slow crossings, such that the light coupling can be supported in a broader range.
In particular, six mode families are highlighted in the map, forming three types of crossings (marked as case-I, -II and -III) where dispersion components could make a decisive role in changing the crossing profile. 
As such, there could be a relative normal, anomalous or close-to-zero dispersion profiles in the auxiliary mode family, with respect to the primary one.

Then, we carried out simulations with qualitatively same dispersion relations.
In details, the dispersion component ($D_{2,p}$) is always fixed to support the soliton comb generation in  the primary mode family.
the absolute value of $D_{2,aux}$ is also fixed but with inverse signs to model normal and anomalous dispersion profiles, and with an higher order component $D_{3,aux}$ to model the close-to-zero dispersion profile. 
In simulations, the nonlinear self-phase modulation and cross phase modulation regarding $A_{\rm aux}$ are neglected, i.e. ${g_{\rm aux} = 0}$ and $g_{\rm c} = 0$, provided that the field pattern in the auxiliary mode family is usually much weaker compared with the soliton pattern in the primary mode family.
A complete list of parameters used in the simulations is shown in Tab. \ref{T2}.

\begin{table}[t!]
\centering
\caption{
Parameters for numerical simulations.
}
\begin{tabular}{c|ccc}
    \hline
     & Case-I & Case-II & Case-III \\
    \hline
    $\kappa_{\rm p}$ & \multicolumn{3}{c}{$2\pi\times 350 ~{\rm kHz}$} \\
    $\kappa_{\rm aux}$ & \multicolumn{3}{c}{$2\pi\times 700 ~{\rm kHz}$} \\
    $\kappa_{\rm ex,p}$ & \multicolumn{3}{c}{$2\pi\times 300 ~{\rm kHz}$} \\
    $\kappa_{\rm ex,aux}$ & \multicolumn{3}{c}{$2\pi\times 550 ~{\rm kHz}$} \\
    $\kappa_{\rm c}$ & $0.6 \times \kappa_{\rm p}$ & $3 \times \kappa_{\rm p}$ & $2 \times \kappa_{\rm p}$ \\
    $|s_{\rm in}|^2$ & \multicolumn{3}{c}{$150 ~{\rm mW}$} \\
    $\alpha$ & \multicolumn{3}{c}{$1/3$} \\
    $^{*}~D_{\rm int}(\mu)$ & \multicolumn{3}{c}{$\mu^2D_{\rm 2,p}/2+\mu^3D_{\rm 3,p}/6$} \\
    $D_{\rm 2,p}$ & \multicolumn{3}{c}{$2\pi\times 2 ~{\rm kHz}$} \\
    $D_{\rm 3,p}$ & \multicolumn{3}{c}{$0$} \\
    $^{*}~D_{\rm aux}(\mu)$ & \multicolumn{3}{c}{$-\Delta \omega + (\mu-\mu_0)^2D_{\rm 2,aux}/2+(\mu-\mu_0)^3D_{\rm 3,aux}/6$} \\
    $\Delta \omega$ & \multicolumn{3}{c}{$35 \times \kappa_{\rm p}$} \\
    $\mu_0$ & $0$ &  $0$ &  $10$ \\ 
    $D_{\rm 2,aux}$ & $2\pi\times 200 ~{\rm Hz}$ & $-2\pi\times 200 ~{\rm Hz}$ & $2\pi\times 200 ~{\rm Hz}$\\
    $D_{\rm 3,aux}$ & $0$ & $0$ & $-2\pi\times 160 ~{\rm Hz}$ \\
    $g_{\rm p}$ & \multicolumn{3}{c}{$2\pi\times 10^{-4} ~{\rm Hz}$} \\
    $g_{\rm aux}$ & \multicolumn{3}{c}{$0$} \\
    $g_{\rm c}$ & \multicolumn{3}{c}{$0$} \\
    $\delta_\omega$ & $35 \times \kappa_{\rm p}$ & $32 \times \kappa_{\rm p}$ & $32 \times \kappa_{\rm p} $ \\
    \hline
    \multicolumn{4}{p{0.9\linewidth}}{
    $^{*}$ The integrated dispersion profile is expressed by a polynomial function up to the third order.
    }
   \end{tabular}
   \label{T2}
\end{table}

As a result, the simulated soliton combs show close spectra to our experimentally observed comb spectra in Fig. \ref{F3}(d). 
In particular, if the sign of the dispersion component $D_2$ is flipped in the auxiliary mode family, we observed different transmitted field patterns, as the result of wave interference. 
Physically, if $D_{\rm 2,aux}>0$ (case-I), the generated comb modes in the primary mode family would be on the red detuned side of the auxiliary modes upon certain detuning ($\delta \omega > 0$), and each mode would feature a phase shift that leads to destructive interference in the transmitted power, see Fig. \ref{F4}(a,d). 
This gives rise to a narrowed power peak on the comb spectrum, which is close to the observation of the comb 2 in Fig. \ref{F3}(d).
In contrast, if $D_{\rm 2,aux}<0$ (case-II), the comb modes would be on the blue-detuned side featuring constructive interference in the transmitted power, see Fig. \ref{F4}(b,e). 
This gives rise to a smooth comb spectrum beyond the standard $sech^2$ profile, which is close to the observation of the comb 3 in Fig. \ref{F3}(d). 
In the third dispersion profile, the comb modes feature both the constructive and destructive interference, resulting in a quick change of comb power when the phase shift is reversed, see Fig. \ref{F4}(c,f).
The transmitted comb is similar to the observation of the soliton comb 1 in Fig. \ref{F3}(d).

Therefore, numerical simulations could confirm the comb spectral tailoring effect via intermode interactions.
Moreover, we realized that the missing nonlinear effects in the simulation could be critical, which may lead to instability of the soliton state if the laser frequency is tuned into the MI region of the coupled auxiliary mode in the anomalous dispersion regime.
This would explain e.g. the comb 2 in Fig. \ref{F3}(d) as an unstable state.
Nevertheless, the simulation could not capture a high-efficient soliton comb, nor the Raman lasing (as Raman effects are switched off).

\section{Conclusion}
In conclusion, we have demonstrated soliton comb generation in ultra-high-$Q$ crystalline fluoride resonators.
The home-developed $\rm MgF_2$ resonators have an intrinsic $Q$-factor approaching $10^{10}$, which is on par with the best performances of the WGM-based microresonators.
Cavity dissipative soliton state can be excited in these resonators, both with the conventional pumping scheme regarding an isolated cavity resonance, and via the intermode pumping scheme.
In the latter, comb spectral tailoring effect has been widely observed and studied, and found dependent on the dispersion regime in the coupled auxiliary mode family.
Indeed, given a large number of transverse mode families in the resonator, the presence of slowly crossings between mode families further reveals structured dispersion profiles in the auxiliary mode families, including a relative normal, anomalous, or close-to-zero dispersion, which are decisive in tailoring the comb spectral envelope.
In addition, the Raman lasing was observed in company to the comb generation in our crystalline resonators, which in contrast to previous understandings could be in a low-noise and stabilized state.
A suspect Raman-soliton comb was also observed around the anti-stokes side, rather than being on the stokes side as previously reported in high-$Q$ silica microresonators \cite{yang_stokes_2016, tan_multispecies_2021}.

Overall, our work reveals certain insights of cavity dissipative structures in nonlinear microresonator systems, and would contribute to the understanding of rich even complex soliton dynamics in the presence of dense mode families.
It also opens an alternative approach to access stable soliton microcombs in crystalline resonators where dispersion engineering is challenging.
Moreover, the results would be supplementary to recent advances in the engineering of intermode interactions in photonic integrated platforms, which has enabled properly tailored soliton comb spectra \cite{tikan_emergent_2021} and super efficient soliton combs\cite{helgason_power-efficient_2022}. 


\begin{backmatter}

\bmsection{Funding}
National Key Research and Development Program of China (2020YFA0309400); National Natural Science Foundation of China (11974234); Shanghai Science and Technology Development Foundation (20QA1403500).

\bmsection{Acknowledgments}
We acknowledge funding from National Key Research and Development Project of China, National Natural Science Foundation of China, and Shanghai Science and Technology Development Funds. This work is also supported by 111 Project (D20031) by MoE of China. Preforms of the crystalline resonators are fabricated in the Engineering Technology Training Center at Shanghai University.

\bmsection{Disclosures}
The authors declare no conflicts of interest.

\bmsection{Data Availability Statement}
Data and simulation codes related to this work are available from the corresponding author upon reasonable request.


\end{backmatter}

\bibliography{Ref}

\begin{thebibliography}{10}
\newcommand{\enquote}[1]{``#1''}

\bibitem{herr_temporal_2014}
T.~Herr, V.~Brasch, J.~D. Jost, C.~Y. Wang, N.~M. Kondratiev, M.~L. Gorodetsky,
  and T.~J. Kippenberg, \enquote{Temporal solitons in optical microresonators,}
  {\protect\JournalTitle{Nature Photonics}} \textbf{8}, 145--152 (2014).

\bibitem{brasch_photonic_2016}
V.~Brasch, M.~Geiselmann, T.~Herr, G.~Lihachev, M.~H.~P. Pfeiffer, M.~L.
  Gorodetsky, and T.~J. Kippenberg, \enquote{Photonic chip–based optical
  frequency comb using soliton {Cherenkov} radiation,}
  {\protect\JournalTitle{Science}} \textbf{351}, 357 (2016).

\bibitem{yi_soliton_2015}
X.~Yi, Q.-F. Yang, K.~Y. Yang, M.-G. Suh, and K.~Vahala, \enquote{Soliton
  frequency comb at microwave rates in a high-q silica microresonator,}
  {\protect\JournalTitle{Optica}} \textbf{2}, 1078--1085 (2015).

\bibitem{kippenberg_tobias_j_dissipative_2018}
{Kippenberg Tobias J.}, {Gaeta Alexander L.}, {Lipson Michal}, and {Gorodetsky
  Michael L.}, \enquote{Dissipative {Kerr} solitons in optical
  microresonators,} {\protect\JournalTitle{Science}} \textbf{361}, eaan8083
  (2018).

\bibitem{stern_battery-operated_2018}
B.~Stern, X.~Ji, Y.~Okawachi, A.~L. Gaeta, and M.~Lipson,
  \enquote{Battery-operated integrated frequency comb generator,}
  {\protect\JournalTitle{Nature}} \textbf{562}, 401--405 (2018).

\bibitem{raja_electrically_2019}
A.~S. Raja, A.~S. Voloshin, H.~Guo, S.~E. Agafonova, J.~Liu, A.~S.
  Gorodnitskiy, M.~Karpov, N.~G. Pavlov, E.~Lucas, R.~R. Galiev, A.~E.
  Shitikov, J.~D. Jost, M.~L. Gorodetsky, and T.~J. Kippenberg,
  \enquote{Electrically pumped photonic integrated soliton microcomb,}
  {\protect\JournalTitle{Nature Communications}} \textbf{10}, 680 (2019).

\bibitem{shen_integrated_2020}
B.~Shen, L.~Chang, J.~Liu, H.~Wang, Q.-F. Yang, C.~Xiang, R.~N. Wang, J.~He,
  T.~Liu, W.~Xie, J.~Guo, D.~Kinghorn, L.~Wu, Q.-X. Ji, T.~J. Kippenberg,
  K.~Vahala, and J.~E. Bowers, \enquote{Integrated turnkey soliton microcombs,}
  {\protect\JournalTitle{Nature}} \textbf{582}, 365--369 (2020).

\bibitem{xiang_chao_laser_2021}
{Xiang Chao}, {Liu Junqiu}, {Guo Joel}, {Chang Lin}, {Wang Rui Ning}, {Weng
  Wenle}, {Peters Jonathan}, {Xie Weiqiang}, {Zhang Zeyu}, {Riemensberger
  Johann}, {Selvidge Jennifer}, {Kippenberg Tobias J.}, and {Bowers John E.},
  \enquote{Laser soliton microcombs heterogeneously integrated on silicon,}
  {\protect\JournalTitle{Science}} \textbf{373}, 99--103 (2021).

\bibitem{chang_integrated_2022}
L.~Chang, S.~Liu, and J.~E. Bowers, \enquote{Integrated optical frequency comb
  technologies,} {\protect\JournalTitle{Nature Photonics}} \textbf{16}, 95--108
  (2022).

\bibitem{marin-palomo_microresonator-based_2017}
P.~Marin-Palomo, J.~N. Kemal, M.~Karpov, A.~Kordts, J.~Pfeifle, M.~H.~P.
  Pfeiffer, P.~Trocha, S.~Wolf, V.~Brasch, M.~H. Anderson, R.~Rosenberger,
  K.~Vijayan, W.~Freude, T.~J. Kippenberg, and C.~Koos,
  \enquote{Microresonator-based solitons for massively parallel coherent
  optical communications,} {\protect\JournalTitle{Nature}} \textbf{546},
  274--279 (2017).

\bibitem{fujii_dissipative_2022}
S.~Fujii, S.~Tanaka, T.~Ohtsuka, S.~Kogure, K.~Wada, H.~Kumazaki, S.~Tasaka,
  Y.~Hashimoto, Y.~Kobayashi, T.~Araki, K.~Furusawa, N.~Sekine, S.~Kawanishi,
  and T.~Tanabe, \enquote{Dissipative {Kerr} soliton microcombs for {FEC}-free
  optical communications over 100 channels,} {\protect\JournalTitle{Optics
  Express}} \textbf{30}, 1351--1364 (2022).

\bibitem{suh_soliton_2018}
M.-G. Suh and K.~J. Vahala, \enquote{Soliton microcomb range measurement,}
  {\protect\JournalTitle{Science}} \textbf{359}, 884 (2018).

\bibitem{trocha_ultrafast_2018}
P.~Trocha, M.~Karpov, D.~Ganin, M.~H.~P. Pfeiffer, A.~Kordts, S.~Wolf,
  J.~Krockenberger, P.~Marin-Palomo, C.~Weimann, S.~Randel, W.~Freude, T.~J.
  Kippenberg, and C.~Koos, \enquote{Ultrafast optical ranging using
  microresonator soliton frequency combs,} {\protect\JournalTitle{Science}}
  \textbf{359}, 887 (2018).

\bibitem{wang_long-distance_2020}
J.~Wang, Z.~Lu, W.~Wang, F.~Zhang, J.~Chen, Y.~Wang, J.~Zheng, S.~T. Chu,
  W.~Zhao, B.~E. Little, X.~Qu, and W.~Zhang, \enquote{Long-distance ranging
  with high precision using a soliton microcomb,}
  {\protect\JournalTitle{Photonics Research}} \textbf{8}, 1964--1972 (2020).

\bibitem{riemensberger_massively_2020}
J.~Riemensberger, A.~Lukashchuk, M.~Karpov, W.~Weng, E.~Lucas, J.~Liu, and
  T.~J. Kippenberg, \enquote{Massively parallel coherent laser ranging using a
  soliton microcomb,} {\protect\JournalTitle{Nature}} \textbf{581}, 164--170
  (2020).

\bibitem{liang_high_2015}
W.~Liang, D.~Eliyahu, V.~S. Ilchenko, A.~A. Savchenkov, A.~B. Matsko,
  D.~Seidel, and L.~Maleki, \enquote{High spectral purity kerr frequency comb
  radio frequency photonic oscillator,} {\protect\JournalTitle{Nature
  Communications}} \textbf{6}, 7957 (2015).

\bibitem{lucas_ultralow-noise_2020}
E.~Lucas, P.~Brochard, R.~Bouchand, S.~Schilt, T.~Südmeyer, and T.~J.
  Kippenberg, \enquote{Ultralow-noise photonic microwave synthesis using a
  soliton microcomb-based transfer oscillator,} {\protect\JournalTitle{Nature
  Communications}} \textbf{11}, 374 (2020).

\bibitem{liu_photonic_2020}
J.~Liu, E.~Lucas, A.~S. Raja, J.~He, J.~Riemensberger, R.~N. Wang, M.~Karpov,
  H.~Guo, R.~Bouchand, and T.~J. Kippenberg, \enquote{Photonic microwave
  generation in the {X}- and {K}-band using integrated soliton microcombs,}
  {\protect\JournalTitle{Nature Photonics}} \textbf{14}, 486--491 (2020).

\bibitem{suh_searching_2019}
M.-G. Suh, X.~Yi, Y.-H. Lai, S.~Leifer, I.~S. Grudinin, G.~Vasisht, E.~C.
  Martin, M.~P. Fitzgerald, G.~Doppmann, J.~Wang, D.~Mawet, S.~B. Papp, S.~A.
  Diddams, C.~Beichman, and K.~Vahala, \enquote{Searching for exoplanets using
  a microresonator astrocomb,} {\protect\JournalTitle{Nature Photonics}}
  \textbf{13}, 25--30 (2019).

\bibitem{obrzud_microphotonic_2019}
E.~Obrzud, M.~Rainer, A.~Harutyunyan, M.~H. Anderson, J.~Liu, M.~Geiselmann,
  B.~Chazelas, S.~Kundermann, S.~Lecomte, M.~Cecconi, A.~Ghedina, E.~Molinari,
  F.~Pepe, F.~Wildi, F.~Bouchy, T.~J. Kippenberg, and T.~Herr, \enquote{A
  microphotonic astrocomb,} {\protect\JournalTitle{Nature Photonics}}
  \textbf{13}, 31--35 (2019).

\bibitem{feldmann_parallel_2021}
J.~Feldmann, N.~Youngblood, M.~Karpov, H.~Gehring, X.~Li, M.~Stappers,
  M.~Le~Gallo, X.~Fu, A.~Lukashchuk, A.~S. Raja, J.~Liu, C.~D. Wright,
  A.~Sebastian, T.~J. Kippenberg, W.~H.~P. Pernice, and H.~Bhaskaran,
  \enquote{Parallel convolutional processing using an integrated photonic
  tensor core,} {\protect\JournalTitle{Nature}} \textbf{589}, 52--58 (2021).

\bibitem{xu_11_2021}
X.~Xu, M.~Tan, B.~Corcoran, J.~Wu, A.~Boes, T.~G. Nguyen, S.~T. Chu, B.~E.
  Little, D.~G. Hicks, R.~Morandotti, A.~Mitchell, and D.~J. Moss, \enquote{11
  {TOPS} photonic convolutional accelerator for optical neural networks,}
  {\protect\JournalTitle{Nature}} \textbf{589}, 44--51 (2021).

\bibitem{liu_high-yield_2021}
J.~Liu, G.~Huang, R.~N. Wang, J.~He, A.~S. Raja, T.~Liu, N.~J. Engelsen, and
  T.~J. Kippenberg, \enquote{High-yield, wafer-scale fabrication of
  ultralow-loss, dispersion-engineered silicon nitride photonic circuits,}
  {\protect\JournalTitle{Nat Commun}} \textbf{12}, 2236 (2021).

\bibitem{braginsky1989quality}
V.~Braginsky, M.~Gorodetsky, and V.~Ilchenko, \enquote{Quality-factor and
  nonlinear properties of optical whispering-gallery modes,}
  {\protect\JournalTitle{Physics Letters A}} \textbf{137}, 393--397 (1989).

\bibitem{grudinin_ultra_2006}
I.~S. Grudinin, A.~B. Matsko, A.~A. Savchenkov, D.~Strekalov, V.~S. Ilchenko,
  and L.~Maleki, \enquote{Ultra high {Q} crystalline microcavities,}
  {\protect\JournalTitle{Optics Communications}} \textbf{265}, 33--38 (2006).

\bibitem{savchenkov_optical_2007}
A.~A. Savchenkov, A.~B. Matsko, V.~S. Ilchenko, and L.~Maleki, \enquote{Optical
  resonators with ten million finesse,} {\protect\JournalTitle{Optics Express}}
  \textbf{15}, 6768--6773 (2007).

\bibitem{sprenger_caf2_2010}
B.~Sprenger, H.~G.~L. Schwefel, Z.~H. Lu, S.~Svitlov, and L.~J. Wang,
  \enquote{{CaF2} whispering-gallery-mode-resonator stabilized-narrow-linewidth
  laser,} {\protect\JournalTitle{Optics Letters}} \textbf{35}, 2870--2872
  (2010).

\bibitem{alnis_thermal-noise-limited_2011}
J.~Alnis, A.~Schliesser, C.~Y. Wang, J.~Hofer, T.~J. Kippenberg, and T.~W.
  Hänsch, \enquote{Thermal-noise-limited crystalline whispering-gallery-mode
  resonator for laser stabilization,} {\protect\JournalTitle{Physical Review
  A}} \textbf{84}, 011804 (2011).

\bibitem{liang_ultralow_2015}
W.~Liang, V.~S. Ilchenko, D.~Eliyahu, A.~A. Savchenkov, A.~B. Matsko,
  D.~Seidel, and L.~Maleki, \enquote{Ultralow noise miniature external cavity
  semiconductor laser,} {\protect\JournalTitle{Nature Communications}}
  \textbf{6}, 7371 (2015).

\bibitem{lim_chasing_2017}
J.~Lim, A.~A. Savchenkov, E.~Dale, W.~Liang, D.~Eliyahu, V.~Ilchenko, A.~B.
  Matsko, L.~Maleki, and C.~W. Wong, \enquote{Chasing the thermodynamical noise
  limit in whispering-gallery-mode resonators for ultrastable laser frequency
  stabilization,} {\protect\JournalTitle{Nature Communications}} \textbf{8}, 8
  (2017).

\bibitem{pavlov_soliton_2017}
N.~G. Pavlov, G.~Lihachev, S.~Koptyaev, E.~Lucas, M.~Karpov, N.~M. Kondratiev,
  I.~A. Bilenko, T.~J. Kippenberg, and M.~L. Gorodetsky, \enquote{Soliton dual
  frequency combs in crystalline microresonators,}
  {\protect\JournalTitle{Optics Letters}} \textbf{42}, 514--517 (2017).

\bibitem{liu_low-loss_2018}
G.~Liu, V.~S. Ilchenko, T.~Su, Y.-C. Ling, S.~Feng, K.~Shang, Y.~Zhang,
  W.~Liang, A.~A. Savchenkov, A.~B. Matsko, L.~Maleki, and S.~J. Ben~Yoo,
  \enquote{Low-loss prism-waveguide optical coupling for ultrahigh-{Q}
  low-index monolithic resonators,} {\protect\JournalTitle{Optica}} \textbf{5},
  219--226 (2018).

\bibitem{pavlov_narrow-linewidth_2018}
N.~G. Pavlov, S.~Koptyaev, G.~V. Lihachev, A.~S. Voloshin, A.~S. Gorodnitskiy,
  M.~V. Ryabko, S.~V. Polonsky, and M.~L. Gorodetsky, \enquote{Narrow-linewidth
  lasing and soliton {Kerr} microcombs with ordinary laser diodes,}
  {\protect\JournalTitle{Nature Photonics}} \textbf{12}, 694--698 (2018).

\bibitem{guo_universal_2016}
H.~Guo, M.~Karpov, E.~Lucas, A.~Kordts, M.~H.~P. Pfeiffer, V.~Brasch,
  G.~Lihachev, V.~E. Lobanov, M.~L. Gorodetsky, and T.~J. Kippenberg,
  \enquote{Universal dynamics and deterministic switching of dissipative
  {Kerr} solitons in optical microresonators,} {\protect\JournalTitle{Nature
  Physics}} \textbf{13}, 94 (2016).

\bibitem{lucas_detuning-dependent_2017}
E.~Lucas, H.~Guo, J.~D. Jost, M.~Karpov, and T.~J. Kippenberg,
  \enquote{Detuning-dependent properties and dispersion-induced instabilities
  of temporal dissipative {Kerr} solitons in optical microresonators,}
  {\protect\JournalTitle{Physical Review A}} \textbf{95}, 043822 (2017).

\bibitem{lucas_breathing_2017}
E.~Lucas, M.~Karpov, H.~Guo, M.~L. Gorodetsky, and T.~J. Kippenberg,
  \enquote{Breathing dissipative solitons in optical microresonators,}
  {\protect\JournalTitle{Nature Communications}} \textbf{8}, 736 (2017).

\bibitem{weng_heteronuclear_2020}
W.~Weng, R.~Bouchand, E.~Lucas, E.~Obrzud, T.~Herr, and T.~J. Kippenberg,
  \enquote{Heteronuclear soliton molecules in optical microresonators,}
  {\protect\JournalTitle{Nature Communications}} \textbf{11}, 2402 (2020).

\bibitem{taheri_all-optical_2022}
H.~Taheri, A.~B. Matsko, L.~Maleki, and K.~Sacha, \enquote{All-optical
  dissipative discrete time crystals,} {\protect\JournalTitle{Nat. Commun.}}
  \textbf{13}, 848 (2022).

\bibitem{guo_intermode_2017}
H.~Guo, E.~Lucas, M.~H. Pfeiffer, M.~Karpov, M.~Anderson, J.~Liu,
  M.~Geiselmann, J.~D. Jost, and T.~J. Kippenberg, \enquote{Intermode
  {Breather} {Solitons} in {Optical} {Microresonators},}
  {\protect\JournalTitle{Physical Review X}} \textbf{7}, 041055 (2017).

\bibitem{herr_mode_2014}
T.~Herr, V.~Brasch, J.~D. Jost, I.~Mirgorodskiy, G.~Lihachev, M.~L. Gorodetsky,
  and T.~J. Kippenberg, \enquote{Mode spectrum and temporal soliton formation
  in optical microresonators,} {\protect\JournalTitle{Phys. Rev. Lett.}}
  \textbf{113}, 123901 (2014).

\bibitem{yi_single-mode_2017}
X.~Yi, Q.-F. Yang, X.~Zhang, K.~Y. Yang, X.~Li, and K.~Vahala,
  \enquote{Single-mode dispersive waves and soliton microcomb dynamics,}
  {\protect\JournalTitle{Nature Communications}} \textbf{8}, 14869 (2017).

\bibitem{tikan_emergent_2021}
A.~Tikan, J.~Riemensberger, K.~Komagata, S.~Hönl, M.~Churaev, C.~Skehan,
  H.~Guo, R.~N. Wang, J.~Liu, P.~Seidler, and T.~J. Kippenberg,
  \enquote{Emergent nonlinear phenomena in a driven dissipative photonic
  dimer,} {\protect\JournalTitle{Nature Physics}} pp. 1--7 (2021).

\bibitem{xue_mode-locked_2015}
X.~Xue, Y.~Xuan, Y.~Liu, P.-H. Wang, S.~Chen, J.~Wang, D.~E. Leaird, M.~Qi, and
  A.~M. Weiner, \enquote{Mode-locked dark pulse kerr combs in normal-dispersion
  microresonators,} {\protect\JournalTitle{Nature Photonics}} \textbf{9},
  594--594 (2015).

\bibitem{jang_dynamics_2016}
J.~K. Jang, Y.~Okawachi, M.~Yu, K.~Luke, X.~Ji, M.~Lipson, and A.~L. Gaeta,
  \enquote{Dynamics of mode-coupling-induced microresonator frequency combs in
  normal dispersion,} {\protect\JournalTitle{Opt. Express}} \textbf{24},
  28794--28803 (2016).

\bibitem{helgason_dissipative_2021}
{\'O}.~B. Helgason, F.~R. Arteaga-Sierra, Z.~Ye, K.~Twayana, P.~A. Andrekson,
  M.~Karlsson, J.~Schröder, and {Victor Torres-Company}, \enquote{Dissipative
  solitons in photonic molecules,} {\protect\JournalTitle{Nat. Photonics}}
  \textbf{15}, 305--310 (2021).

\bibitem{xue_super-efficient_2019}
X.~Xue, X.~Zheng, and B.~Zhou, \enquote{Super-efficient temporal solitons in
  mutually coupled optical cavities,} {\protect\JournalTitle{Nat. Photonics}}
  \textbf{13}, 616--622 (2019).

\bibitem{helgason_power-efficient_2022}
{\'O}.~B. Helgason, M.~Girardi, Z.~Ye, F.~Lei, J.~Schröder, and V.~T. Company,
  \enquote{Power-efficient soliton microcombs,}
  {\protect\JournalTitle{{arXiv}:2202.09410 [physics]}}  (2022).

\bibitem{li_stably_2017}
Q.~Li, T.~C. Briles, D.~A. Westly, T.~E. Drake, J.~R. Stone, B.~R. Ilic, S.~A.
  Diddams, S.~B. Papp, and K.~Srinivasan, \enquote{Stably accessing
  octave-spanning microresonator frequency combs in the soliton regime,}
  {\protect\JournalTitle{Optica}} \textbf{4}, 193--203 (2017).

\bibitem{weng_directly_2021}
H.~Weng, J.~Liu, A.~A. Afridi, J.~Li, J.~Dai, X.~Ma, Y.~Zhang, Q.~Lu, J.~F.
  Donegan, and W.~Guo, \enquote{Directly accessing octave-spanning dissipative
  {Kerr} soliton frequency combs in an {AlN} microresonator,}
  {\protect\JournalTitle{Photonics Research}} \textbf{9}, 1351--1357 (2021).

\bibitem{lei_thermal_2022}
F.~Lei, Z.~Ye, and V.~Torres-Company, \enquote{Thermal noise reduction in
  soliton microcombs via laser self-cooling,} {\protect\JournalTitle{Opt.
  Lett.}} \textbf{47}, 513--516 (2022). Publisher: OSA.

\bibitem{chembo_spatiotemporal_2015}
Y.~K. Chembo, I.~S. Grudinin, and N.~Yu, \enquote{Spatiotemporal dynamics of
  {Kerr}-{Raman} optical frequency combs,} {\protect\JournalTitle{Physical
  Review A}} \textbf{92} (2015).

\bibitem{cherenkov_raman-kerr_2017}
A.~V. Cherenkov, N.~M. Kondratiev, V.~E. Lobanov, A.~E. Shitikov, D.~V.
  Skryabin, and M.~L. Gorodetsky, \enquote{Raman-kerr frequency combs in
  microresonators with normal dispersion,} {\protect\JournalTitle{Optics
  Express}} \textbf{25}, 31148--31158 (2017).

\bibitem{yu_raman_2020}
M.~Yu, Y.~Okawachi, R.~Cheng, C.~Wang, M.~Zhang, A.~L. Gaeta, and M.~Lončar,
  \enquote{Raman lasing and soliton mode-locking in lithium niobate
  microresonators,} {\protect\JournalTitle{Light: Science \& Applications}}
  \textbf{9}, 9 (2020).

\bibitem{xia_engineered_nodate}
D.~Xia, Y.~Huang, B.~Zhang, P.~Zeng, J.~Zhao, Z.~Yang, S.~Sun, L.~Luo, G.~Hu,
  D.~Liu, Z.~Wang, Y.~Li, H.~Guo, and Z.~Li, \enquote{Engineered raman lasing
  in photonic integrated chalcogenide microresonators,}
  {\protect\JournalTitle{Laser \& Photonics Reviews}} \textbf{n/a}, 2100443
  (2022).

\bibitem{yang_stokes_2016}
Q.-F. Yang, X.~Yi, K.~Y. Yang, and K.~Vahala, \enquote{Stokes solitons in
  optical microcavities,} {\protect\JournalTitle{Nature Physics}} \textbf{13},
  53 (2016).

\bibitem{tan_multispecies_2021}
T.~Tan, Z.~Yuan, H.~Zhang, G.~Yan, S.~Zhou, N.~An, B.~Peng, G.~Soavi, Y.~Rao,
  and B.~Yao, \enquote{Multispecies and individual gas molecule detection using
  stokes solitons in a graphene over-modal microresonator,}
  {\protect\JournalTitle{Nat Commun}} \textbf{12}, 6716 (2021).

\bibitem{coillet_microwave_2013}
A.~Coillet, R.~Henriet, H.~Kien~Phan, M.~Jacquot, L.~Furfaro, I.~Balakireva,
  L.~Larger, and Y.~K. Chembo, \enquote{Microwave {Photonics} {Systems} {Based}
  on {Whispering}-gallery-mode {Resonators},}
  {\protect\JournalTitle{Jove-Journal Of Visualized Experiments}}  (2013).

\bibitem{lin_barium_2014}
G.~Lin, S.~Diallo, R.~Henriet, M.~Jacquot, and Y.~K. Chembo, \enquote{Barium
  fluoride whispering-gallery-mode disk-resonator with one billion
  quality-factor,} {\protect\JournalTitle{Optics Letters}} \textbf{39},
  6009--6012 (2014).

\bibitem{fujii_all-precision-machining_2020}
S.~Fujii, Y.~Hayama, K.~Imamura, H.~Kumazaki, Y.~Kakinuma, and T.~Tanabe,
  \enquote{All-precision-machining fabrication of ultrahigh-{Q} crystalline
  optical microresonators,} {\protect\JournalTitle{Optica}} \textbf{7},
  694--701 (2020).

\bibitem{dumeige_determination_2008}
Y.~Dumeige, S.~Trebaol, L.~Ghişa, T.~K.~N. Nguyên, H.~Tavernier, and
  P.~Féron, \enquote{Determination of coupling regime of high-{Q} resonators
  and optical gain of highly selective amplifiers,}
  {\protect\JournalTitle{Journal of the Optical Society of America B}}
  \textbf{25}, 2073--2080 (2008).

\bibitem{lecaplain_mid-infrared_2016}
C.~Lecaplain, C.~Javerzac-Galy, M.~L. Gorodetsky, and T.~J. Kippenberg,
  \enquote{Mid-infrared ultra-high-{Q} resonators based on fluoride crystalline
  materials,} {\protect\JournalTitle{Nature Communications}} \textbf{7}, 13383
  (2016).

\bibitem{spillane_ideality_2003}
S.~M. Spillane, T.~J. Kippenberg, O.~J. Painter, and K.~J. Vahala,
  \enquote{Ideality in a {Fiber}-{Taper}-{Coupled} {Microresonator} {System}
  for {Application} to {Cavity} {Quantum} {Electrodynamics},}
  {\protect\JournalTitle{Physical Review Letters}} \textbf{91}, 043902 (2003).

\bibitem{pfeiffer_coupling_2017}
M.~H. Pfeiffer, J.~Liu, M.~Geiselmann, and T.~J. Kippenberg, \enquote{Coupling
  {Ideality} of {Integrated} {Planar} {High}-{Q} {Microresonators},}
  {\protect\JournalTitle{Physical Review Applied}} \textbf{7}, 024026 (2017).

\bibitem{neelamraju_experimental_2012}
S.~Neelamraju, A.~Bach, J.~C. Schön, D.~Fischer, and M.~Jansen,
  \enquote{Experimental and theoretical study on {Raman} spectra of magnesium
  fluoride clusters and solids,} {\protect\JournalTitle{The Journal of Chemical
  Physics}} \textbf{137}, 194319 (2012).

\bibitem{xue_microresonator_2017}
X.~Xue, P.-H. Wang, Y.~Xuan, M.~Qi, and A.~M. Weiner, \enquote{Microresonator
  {Kerr} frequency combs with high conversion efficiency,}
  {\protect\JournalTitle{Laser \& Photonics Reviews}} \textbf{11},
  1600276--1600276 (2017).

\bibitem{marin-palomo_performance_2020}
P.~Marin-Palomo, J.~N. Kemal, T.~J. Kippenberg, W.~Freude, S.~Randel, and
  C.~Koos, \enquote{Performance of chip-scale optical frequency comb generators
  in coherent {WDM} communications,} {\protect\JournalTitle{Optics Express}}
  \textbf{28}, 12897--12910 (2020). Publisher: OSA.

\end{thebibliography}


\end{document}